\documentclass[11pt]{article}
\usepackage{amsfonts}
\usepackage{latexsym}
\title{ New solution of $D=11$ supergravity on $S^7$ from $D=4$}
\author{H\'ector H. Hern\'andez Hern\'andez\footnote{e-mail: hehh@xanum.uam.mx.} \\ and
\and Hugo A. Morales T\'ecotl\footnote{e-mail: hugo@xanum.uam.mx.
Associated member of ICTP, Trieste Italy.} \\
{\small Departamento de F\'{\i}sica, Universidad Aut\'onoma
Metropolitana Iztapalapa}\\ {\small A.P. 55-534, M\'exico D.F.
09340, M\'exico}}
\date{}

\begin{document}
\maketitle
\begin{abstract}
A new static partially twisted solution of $N=4$, $SO(4)$ gauged
supergravity in $D=11$ is obtained in this work using Cveti\^c et
al embedding of four dimensional into eleven dimensional
supergravities. In four dimensions we get two solutions: an
asymptotic one corresponding to $AdS_4$ and a near horizon fixed
point solution of the form $AdS_2\times H_2$. Hence, while the
former solution has 32 supercharges the latter turns out to have
only 4 conserved. Moreover, we managed to find an exact
interpolating solution, thus connecting the above two. Aiming at a
future study of $AdS/CFT$ duality for the theory at hand we
derived the Penrose limit of the four dimensional solutions.
Interestingly the pp-wave limit of the near horizon solution
suggests itself as being of the supernumerary supersymmetric type.
In $D=11$ we exhibit the uplift of the four dimensional solutions:
one associated to $AdS_4\times S^7$ and the other to a foliation
of $AdS_2\times H_2 \times S^7$, as well as their pp-wave limits.
\end{abstract}

\section{Introduction.}
Within the current form of string theory it is commonly accepted
there exists a duality between superconformal field theories with
large $N$ and weakly coupled supergravities on $AdS$ backgrounds
\cite{malda1,gubser,witten}.  Also non conformal theories have been recently studied
\cite{malda2}. Whenever there are solutions to supergravity
theories preserving part of the original supersymmetry they might
provide new evidence of the $AdS/CFT$ duality establishing a
correspondence with twisted SCYM theories. The twist term having
origin in the projections imposed upon the Killing spinors
\cite{super1,super2}.

One way to look for $AdS/CFT$ dualities is to obtain
solutions from the supergravity
side wrapping or twisting known BPS brane solutions of eleven or
ten dimensional supergravities \cite{Nunez1}. The solutions are determined
using the first order supersymmetry equations to obtain
the metric and field configurations and this tells us about the
supersymmetry (partial or total) of the solution.

Alternatively one can uplift lower dimensional supergravity
solutions to eleven or ten dimensions \cite{Cvetic1,
Cvetic2,Cvetic3} which correspond, then, to a twisted or partially
twisted supergravity in eleven or ten dimensions, respectively.

In this work we perform the uplifting procedure from $D=4$ to
$N=4,SO(4)$ gauged supergravity in $D=11$ \cite{Cvetic1}:
we use a metric ansatz
for the four dimensional metric of the form $AdS_2\times H_2$,
where $H_2$ corresponds to a two dimensional hyperbolic space.

Starting from the uplift formulae in \cite{Cvetic1} we determined
two solutions of the four dimensional supergravity. One is the
vacuum  $AdS_4$ with all matter fields set to zero which preserves
the whole supersymmetry (32 supercharges). This admits an
interpretation as describing an asymptotic behavior. The other
solution has as matter contents two copies of $SU(2)$ gauge fields
pointing in one of the three $S^3$ directions. The two $SU(2)$
gauge couplings being equal, and the scalar fields (dilaton and
axion) set to zero. In this case there is the number of 4
supercharges conserved. This might be related to the equality of
the coupling constants and the way the brane wraps onto the
transverse $S^7$ \cite{super1,super2}. Here the interpretation
corresponds to a near horizon regime since it includes the $ r=0$
region corresponding to the $AdS_2$ horizon. Interestingly we were
able to find an exact interpolating solution connecting the above
two.

As for the $AdS_2\times H_2$ solution, when we approach the
$AdS_2$ horizon, such a factor geometry dominates over the $H_2$.
A low energy effective theory is obtained in $0+1$ dimensions
which can be interpreted as a Super Conformal Quantum Mechanics
that can be expressed in terms of $M2$ and $M5$ branes from the
whole theory \cite{stromin1, stromin2,stromin3}. This is of
interest as a way of studying (multi) black hole(s) configurations
and the analysis of their quantum behavior .

Similar analysis to the one we present in this work have been
made to find other supersymmetric solutions with diverse matter contents
(see \cite{sols1} and references therein).

In connection with $AdS/CFT$ duality the parallel plane wave
(Penrose) limits of supergravity scenarios have received much
attention recently \cite{berenstein,penrose, guven,correa1,nunez2}.
Such a limit could define a Matrix model to check the relation between
the gravity and the gauge theory side \cite{nastase} --which we do not pursue
here. With this motivation we have obtained the Penrose limit of
our solutions in $D=4$ and determined that one of them
exhibits a supernumerary supersymmetry
\cite{super1,super2,super3}. It presents an enhancement of the original
preserved supersymmetry. Such behavior has been observed
in previous work \cite{correa1}.

This paper is organized as follows. In section 2 we provide the
details in determining the two solutions in four dimensional
supergravity. The uplifting to eleven dimensions is included.
Section 3 is devoted to show the amount of supersymmetry preserved
by the four dimensional solutions and we use this technic to
obtain an exact interpolating solution between those obtained in
the previous section. In section 4 the pp-wave limit of our
solutions are obtained together with the amount of supersymmetry
they preserve following an analogous analysis as in section 3.
Finally section 5 contains a discussion of our results.

\section{New solutions in $D=4$ and $D=11$.}
In this section we obtain a bosonic solution for the $SO(4),N=4$
four dimensional supergravity described in \cite{Cvetic1} as
follows. Since we would like to know what dual description such a solution
admits
in terms of a super conformal Yang-Mills theory
we propose a background containing an $AdS$ factor, so we begin with a
metric ansatz of the form
\begin{equation}\label{ansatz}
ds_4^2=f(r)ds_{AdS_2}^2+g(r)ds_{H_2}^2, \end{equation} where $H_2$
is a two dimensional hyperbolic manifold and $r$ is the radial
coordinate of $AdS_2$.

A particular form of the $SU(2)$ gauge fields with equal coupling
constants and vanishing scalar fields are plugged in the resulting
Einstein-Yang-Mills field equations. Since we want the solution to
preserve some fraction of supersymmetry we look for field
configurations satisfying the corresponding first order
supersymmetry conditions (\cite{Das1}). We succeeded in doing so
by introducing certain Killing spinor projections.

We found two solutions. One corresponds to a near horizon
configuration
of the four dimensional theory, whereas the other can be
read as representing an asymptotic behavior with all matter fields turned off.

Let us start with the four dimensional bosonic Lagrangian for $SO(4),
N=4$ gauged supergravity \cite{Cvetic1}
\begin{eqnarray}\label{eq:lagrange}
\nonumber \mathcal{L}& =&
R-\frac{1}{2}(\partial_{\mu}\phi)^2-\frac{1}{2}e^{2\phi}(\partial_{\mu}\chi)^2+2\alpha^2(4+2\cosh
\phi +\chi^2 e^{\phi})-\frac{1}{2}e^{-\phi}F^2\\
&-&\frac{e^{\phi}}{2(1+\chi^2 e^{2\phi})}\tilde{F}^2 -
\frac{\chi}{2}F \ast F+\frac{\chi e^{2\phi}}{2(1+ \chi^2
e^{2\phi})}\tilde{F}\ast \tilde{F}.
\end{eqnarray}
$A,\tilde{A}$ are $SU(2)$ gauge fields with field strengths
$F^{i}=dA^{i}+\frac{1}{2}\alpha \varepsilon^{ijk}A^{j}A^{k},$ and
similarly for $\tilde{F}$. Here we have set the two coupling constants equal
$g=\tilde{g}=\alpha$.  $\phi,\chi$ are scalar fields (dilaton and
axion respectively).

From this we can write the Einstein equations
\begin{eqnarray}
\nonumber R_{\mu\nu} &=& \frac{1}{2}\partial_{\mu}\phi \partial_{\nu}\phi+\frac{e^{2\phi}}{2}\partial_{\mu}\chi \partial_{\nu}\chi+ e^{-\phi}\left( F_{\mu\alpha}F_{\nu}^{\alpha}- \frac{g_{\mu\nu}}{4}F^2 \right) \\
 &+& \frac{e^{\phi}}{1+\chi^2e^{2\phi}}\left( \tilde{F}_{\mu\alpha}\tilde{F}_{\nu}^{\alpha}- \frac{g_{\mu\nu}}{4}\tilde{F}^2 \right)-6\alpha^2g_{\mu\nu}.
\end{eqnarray}

Now we asume $A_{\mu}=\tilde{A}_{\mu}$. As for the YM equations of
motion we get
\begin{equation}\label{eq:YM}
\frac{1}{\sqrt{-g}}\partial_{\mu}\left(\sqrt{-g}e^{-\phi}F^{\mu\nu a}\right)=\left(A_{\mu}^b F^{\mu\nu c}-\frac{\chi}{\sqrt{-g}}\varepsilon^{\mu\nu\alpha\beta}A_{\mu}^b F_{\alpha\beta}^c\right)\varepsilon^{abc}.
\end{equation}

Our ansatz consists of setting $\phi=\chi=0$, together with the
metric (\ref{ansatz}):
\begin{equation}\label{ansatz1}
ds^2=e^{2f(r)}\left( -dt^2+dr^2\right)
+\frac{e^{2h(r)}}{y^2}\left( dx^2+dy^2\right),
\end{equation} where $f(r),h(r)$ are functions to be determined.

A direct calculation using the YM equations (\ref{eq:YM}) yields
the following field solution
\begin{equation}\label{eq:A-fields}
A_{x}^{3}=\tilde{A}_{x}^{3}=\frac{k}{y} \Rightarrow F_{xy}^{3}=
\tilde{F}_{xy}^3=\frac{k}{y^2}; \quad \phi=\chi=0,
\end{equation}
with $k$ a constant to be determined later. Next we write the
components of the Einstein equations for this configuration
\begin{eqnarray}
R_{11}&=& f''+2f' h',\\
R_{22}&=& -f''-2(h')^2-2h''+2f'h', \\
R_{33}=R_{44}&=& -\frac{1}{y^2}\left(
2e^{2(h-f)}(h')^2+e^{2(h-f)}h''+1\right).
\end{eqnarray}

Now we let $h(r)=\textrm{const}$ (fixed point) and find
\begin{eqnarray}
\label{eh}e^{-2h}&=&\frac{-1+ \sqrt{1+24(k \alpha)^2}}{2k^2},\\
\label{ef}e^{2f(r)}&=&\frac{2}{F^2+12\alpha^2}\frac{1}{r^2},
\end{eqnarray}
where $F^2=2k^2e^{-4h}$.
In light of (\ref{eq:sol}) we obtain the metric corresponding to
$AdS_2 \times H_2$. Due to the fact that the $AdS_2$-horizon lies
on $r=0$, the near horizon region is included within this description.

It is worth noticing that the matter free solution
$A=\tilde{A}=0$ to the above equations reads
\begin{equation}\label{AdS4}
ds^2=\frac{2}{\alpha^2 r^2}(-dt^2+dr^2+dx^2+dy^2),
\end{equation}
just $AdS_4$. Let us observe that (\ref{eq:A-fields}) yields zero for
$y\rightarrow \infty$ and since (\ref{AdS4}) corresponds
to a matter free configuration, we refer to it as asymptotic.

Now we proceed to the uplifting from four to eleven dimensions \cite{Cvetic1}.
We get
\begin{enumerate}
\item $AdS_4 \times S^7$:
\begin{equation}\label{eq:vacuum}
ds_{11}^2=\frac{2}{\alpha^2
r^2}(-dt^2+dr^2+dx^2+dy^2)+\frac{2}{\alpha^2}
d\xi^2+\frac{1}{2\alpha^2}\left[\cos^2\xi d\Omega_3+\sin^2 \xi
d\tilde{\Omega}_3 \right],
\end{equation}
where $d\Omega_3$ is the metric on the three sphere.

\item $AdS_2\times H_2 \times \tilde{S}^7$, tilde meaning the squashing
of $S^7$
\begin{eqnarray}
\label{eq:sol}\lefteqn{ds_{11}^2=\frac{2}{F^2+12\alpha^2}\frac{1}{r^2}\left(-dt^2+dr^2\right)
+\frac{e^{2h}}{y^2}\left(dx^2+dy^2\right)+\frac{2}{\alpha^2}d\xi^2+{}} \nonumber\\
& &{}+\frac{1}{2\alpha^2}\left[\cos^2\xi \left(
\sigma_{1}^2+\sigma_{2}^2+\left[
\sigma_{3}-\frac{dx}{y}\right]^2\right)+\sin^2 \xi
\left(\sigma\leftrightarrow \tilde{\sigma}\right) \right].
\end{eqnarray}
\end{enumerate}
In this last equation $\sigma^i$ are left invariant one forms\footnote{
\begin{eqnarray}\sigma^1&=&
d\rho+\cos{\zeta}d\tau,\\ \sigma^2&=& \cos{\rho}d\zeta+\sin{\zeta}
\sin{\rho}d\tau,\\
\sigma^3&=&\sin{\rho}d\zeta-\sin{\zeta}\cos{\rho}d\tau.
\end{eqnarray}} in
$SO(3)\simeq SU(2)$.

For completeness we write down the gauge and strength fields.
In four dimensions we have
\begin{eqnarray}\label{eq:fields}
A^{i}=\tilde{A}^{i}&=&\frac{k}{y}dx \ \delta^{i}_3,\\
F^{i}=\tilde{F}^{i}&=&\frac{k}{y^2}dx\wedge dy \ \delta^{i}_3.
\end{eqnarray}
In the case of eleven dimensions we adopt the one forms of \cite{Cvetic1}
\begin{equation}
h^i=\sigma^i-\alpha A^i,
\end{equation}
where $A^i$ is given in (\ref{eq:fields}),
and similarly for $\tilde{h}^i$. Then the eleven dimensional four
form field strength reads
\begin{eqnarray}
F_{(4)}&=& -3\sqrt{2}\alpha \varepsilon_{(4)}+F'_{(4)},\nonumber\\
\sqrt{2}\alpha^2 F'_{(4)}&=&\sin{\xi}^2 \cos{\xi}^2 d\xi \wedge
\left(h^i\wedge F^i-\bar{h}^i\wedge \bar{F}^i\right)+ \nonumber \\
& & +\frac{1}{4}\epsilon_{ijk}\left( \cos{\xi}^2 h^i\wedge h^j
\wedge *F^k+ \sin{\xi}^2 \bar{h}^i\wedge \bar{h}^j \wedge
*\bar{F}^k \right).
\end{eqnarray}

The near horizon and asymptotic character of our solutions
suggests to look for a metric that connects smoothly between the
two. Since we will use the supersymmetry transformation equations
we postpone this analysis until the next section. Also we want to
determine what is the fraction of supersymmetry retained by this
solution. We pursue this in the next section.

\section{Supersymmetry of the solutions.}
As for our $AdS_4$  solution it is well known it is maximally
supersymmetric \cite{freedman,lu1,lu2,Romans1,blau2}, namely it
has 32 supercharges.

Regarding the near horizon solution (\ref{eq:sol}) we now
determine the amount of supersymmetry preserved. To this end we
use the supersymmetry transformations
\cite{Das1,Das2,Barton1,barton2}
\begin{eqnarray}
\label{eq:susy1}\delta \bar{\chi}^{i}&=&\frac{1}{2\sqrt{2}}\varepsilon^{ijkl}
\bar{\epsilon}^{j}\sigma^{\mu\nu}F_{\mu\nu}^{kl}=0,\\
\label{eq:susy2} \delta\bar{\psi}_{\lambda}^{i}&=& \bar{\epsilon}^{i}
\overleftarrow{D}_{\lambda}-\frac{1}{2}\bar{\epsilon}^{j}\gamma_{\lambda}
\sigma^{\mu\nu}F_{\mu\nu}^{ij}-\alpha\bar{\epsilon}^{i}\gamma_{\lambda}=0,
\end{eqnarray}
where $D_{\mu}\psi_{\nu}^{I}=\left(
\partial_{\mu}+\frac{1}{2}\omega_{\mu
ab}\sigma^{ab}\right)\psi_{\nu}^{i}+2\alpha
A_{\mu}^{ij}\psi_{\nu}^{j}$, and $\omega_{a\mu\nu}$ is the spin
connection. Our conventions are
\begin{equation}
 \sigma^{ab}=\frac{1}{4}[\gamma^{a},\gamma^{b}], \quad \gamma^{\mu}=e_{a}^{\mu}\gamma^{a},
\end{equation}
and
\label{eq:gamma}\[ \gamma^0=i\left( \begin{array}{cc}
0 & \sigma^2\\
\sigma^2 & 0 \end{array} \right), \quad
\gamma^1=-\left( \begin{array}{cc}
\sigma^3 & 0\\
0& \sigma^3 \end{array} \right),\]
\begin{equation} \gamma^2=i\left( \begin{array}{cc}
0 & -\sigma^2\\
\sigma^2 & 0 \end{array} \right), \quad
\gamma^3=\left( \begin{array}{cc}
\sigma^1 & 0\\
0& \sigma^1 \end{array} \right).
\end{equation}
Here latin indices are flat index and greek indices are curved.

We also need the spin connection for (\ref{ansatz1}). We get
\begin{eqnarray}
\omega_{t\hat{t}\hat{r}}&=&\frac{1}{r},\\
\omega_{x\hat{y}\hat{x}}&=&\frac{1}{y}.
\end{eqnarray}
Combining (\ref{eq:fields}) and (\ref{eq:susy1}) we have
$\varepsilon^{1}=\varepsilon^{4}=0$.

Now we write eq. (\ref{eq:susy2}) in components
\begin{eqnarray}
\label{eq:epst}\bar{\varepsilon}^2\left(\overleftarrow{\partial}_t+
\frac{1}{r}\sigma^{tr}+\alpha e^{f} \gamma^{t}\right)+\bar{\varepsilon}^3
\left(ke^{f-2h} \gamma^{t}\sigma^{xy}\right)&=&0,\\
\bar{\varepsilon}^2\left(\overleftarrow{\partial}_{r}-\alpha e^f\gamma^{r}
\right)-\bar{\varepsilon}^3\left(ke^{f-2h} \gamma^{r}\sigma^{xy}\right)&=&0,\\
\bar{\varepsilon}^2\left(\overleftarrow{\partial}_{x}+\frac{1}{y}\sigma^{yx}-
\frac{\alpha e^h}{y}\gamma^{x}\right)+\bar{\varepsilon}^3\left(
\frac{2\alpha k}{y}-\frac{k}{e^h y} \gamma^{x}\sigma^{xy}\right)&=&0,\\
\bar{\varepsilon}^2\left(\overleftarrow{\partial}_{y}-\frac{\alpha e^h}{y}
\gamma^{y}\right)-\bar{\varepsilon}^3\left(\frac{k}{e^h y} \gamma^{y}
\sigma^{xy}\right)&=&0,\label{eq:epsy}
\end{eqnarray}

Imposing $\partial_\mu \bar{\varepsilon}^2=\partial_\mu
\bar{\varepsilon}^3=0,\quad \mu=t,x,y$ on the spinors as in
\cite{sabra}, the projections, with $i=2,3$,
\begin{eqnarray}
\label{proy1}\bar{\varepsilon}^i \gamma^x \gamma^y & = & i\bar{\varepsilon}^{i}, \\
\bar{\varepsilon}^i \gamma^r & = & \bar{\varepsilon}^{i},
\end{eqnarray}
and substituting everything in (\ref{eq:epst}) we finally end up with
\begin{eqnarray}
\label{spinors23}\bar{\varepsilon}^2&=& i \bar{\varepsilon}^3,\\
\bar{\varepsilon}^2(r)&=&\sqrt{r}\bar{\varepsilon}_0,
\end{eqnarray} which solve
(\ref{eq:susy1},\ref{eq:susy2}) with metric (\ref{ansatz1}) and (\ref{eh},\ref{ef}).

The projection (\ref{proy1}) together with (\ref{spinors23})
reduce the number of independent spinor components of
$\epsilon^2,\epsilon^3$  to 4. Using the triviality of
$\epsilon^1$ and $\epsilon^4$ we conclude that the fraction of
supersymmetry preserved by the near horizon solution is 1/8.

This corresponds to 4 supercharges retained by this supergravity
solution in four dimensions. Remarkably such a number of
supercharges is not commonly found. Let us stress this number
stems from the twisting given by the projection (\ref{proy1}).

Given our near horizon solution $AdS_2 \times H_2$ in four
dimensions it is interesting to see what happens in the IR: as we
lower the energy of the massless modes their wave length becomes
large compared to the size of the $(x,y)$ coordinates in the
transverse space and there is no dependence in those coordinates;
so we end up with an effectively $0+1$ dimensional field theory.
So, the supergravity theory at hand is dual to a $2+1$ super
conformal Yang-Mills theory and at low energies is dual to a Super
Conformal Quantum Mechanics with four supercharges (see
\cite{stromin1,stromin2,stromin3}) whose field description may be
given in terms of the M2brane of the full theory.

Next we compute the interpolating solution we mentioned at the end
of the previous section. To this end we use again the ansatz
(\ref{ansatz})
\begin{equation}
ds^2=e^{2f(r)}\left( -dt^2+dr^2 \right)+
\frac{e^{2h(r)}}{y^2}\left( dx^2+dy^2\right).
\end{equation}

As the Einstein equations are second order and non linear, and we
want the solution to preserve some supersymmetry we impose the
supersymmetry first order condition to the above ansatz. We use
the formulae given above.

The non zero spin connection components for this metric are
\begin{eqnarray}
\omega_{t01}&=&f',\\
\omega_{x12}=\omega_{y13}&=& e^{h-f}\frac{h'}{y},\\
\omega_{x23}&=& \frac{1}{y},
\end{eqnarray}
where the prime denotes derivative with respect to $r$. From
(\ref{eq:susy2}) the supersymmetry condition now reads\footnote{We
use $M^{ij}=\delta_{2}^{[i}\delta_{3}^{j]}$.}
\begin{eqnarray}
0&=&\bar{\epsilon}^{i}\left( f'\sigma^{01}+\alpha
e^{f}\gamma^0\right)+\bar{\epsilon}^{j}
\left( -\sigma^{23}\gamma^0 k e^{-2h+f}\right)M^{ij},\\
0&=&\bar{\epsilon}^{i}\left( e^{h-f}h'
\sigma^{12}+\sigma^{23}-\alpha e^{h}
\gamma^2\right) +\bar{\epsilon}^{j}\left( 2\alpha k+\sigma^{23}\gamma^2 ke^{-h}\right)M^{ij},\\
0&=&\bar{\epsilon}^{i}\left( e^{h-f}h' \sigma^{13}-\alpha e^h
\gamma^3\right) + \bar{\epsilon}^{j} \left( \sigma^{23}\gamma^3 k
e^{-h}\right)M^{ij},
\end{eqnarray}
plus an equation yielding the radial dependence of the spinor.

We can use the projection
$\bar{\epsilon}^{j}\sigma^{23}M^{ij}=\bar{\epsilon}^{i}$.
Substituting back in the last system we obtain from the first and
third equations $h'(r)=f'(r)$. If we use $h(r)=f(r)$ and the
second equation we obtain that $\alpha\propto k$ and
\begin{equation}\label{eq:interpolante}
f(r)=h(r)=\ln \frac{\tanh \left( r+C_1\right)}{C_2\alpha}.
\end{equation}

We have obtained a near horizon and an asymptotic solution which
is interpolated by (\ref{eq:interpolante}) in four dimensions,
that corresponds to a M2 brane solution to M theory.

\section{Penrose limit in $D=4$ and $D=11$.}
In this section we obtain the Penrose limit of our four
dimensional solutions
(\ref{ansatz1},\ref{eh},\ref{ef},\ref{AdS4}) that corresponds to pp-wave
configurations. We include also the uplifting from four to eleven dimensions.
In the sequel we follow \cite{correa1,blau1,fuji,guven2}.

Let us start with the near horizon solution and write it down
as
\begin{equation}
ds^2= e^{2f(r)}\left( -dt^2+dr^2 \right)+e^{2h}\left( d\theta^2
+\sinh{\theta}^2 d\phi ^2 \right).
\end{equation}

Since we want to make the limit on a null geodesic on $\theta= {\rm
sinh}^{-1}1,\quad \phi=0$ we introduce new coordinates $(u,v,p,q)$ by
\begin{eqnarray}
t&=& \frac{1}{2}\Omega \left( u-\Omega^2 v\right),\\
r&=& \frac{1}{2}\Omega \left( u+\Omega^2 v\right),\\
\theta &=& \Omega p+{\rm
sinh}^{-1}1,\\
\phi&=& \Omega q.
\end{eqnarray}
These coordinates are motivated by the coupling-dependent
factor in (\ref{ansatz1},\ref{eh},\ref{ef}). We need to redefine the gauge fields so
that all the terms in the Lagrangian (\ref{eq:lagrange}) acquire
the same factor of $\Omega$. This can be done redefinition of the following fields
\begin{eqnarray}
\bar{g}_{\mu\nu}&=& \Omega^{-2}g_{\mu\nu},\\
\bar{A}&=& \Omega^{-1}A,\\
\bar{\alpha}&=&\Omega \alpha.
\end{eqnarray}
After performing the limit $\Omega\rightarrow 0$ the metric and the
fields read
\begin{eqnarray}\label{ppAdS2H2}
d \bar{s}^2&=&2k^2 \left[ \frac{1}{u^2}\left( dudv \right)
+\left(dp^2+dq^2\right) \right],\\
\bar{A}^3&=&\frac{k}{\sqrt{2}}(dp+dq),\\
\label{eq:limitfield}\bar{F}^3&=&0.
\end{eqnarray}

In order to make the uplifting to 11 dimensions of this solution
we need to make a similar rescaling on the remaining seven
coordinates of the transverse space as follows
\begin{equation}
\xi \rightarrow \Omega^2\xi+\frac{\pi}{4},
\end{equation}
and homogeneous transformation for the angular coordinates
($\rho,\zeta,\tau$) parameterizing the left invariant one-forms.
Altogether we obtain \\
\begin{eqnarray}\label{ppeleven}
\nonumber d\bar{s}^2 &=& 2k^2\left[ \frac{1}{u^2}\left( dudv
\right) +2\left(dp^2+dq^2\right)\right]+\\ \nonumber &&
\frac{1}{4\alpha^2}\Big[ 2d\xi^2+ \left(
d\rho^2+d\zeta^2+d\tau^2+2d\zeta d\tau\right)+\\ &&\left(
d\tilde{\rho}^2+d\tilde{\zeta}^2+d\tilde{\tau}^2+2d\tilde{\zeta}
d\tilde{\tau}\right)\Big].
\end{eqnarray}

Next we take the Penrose limit of the asymptotic solution
(\ref{eq:vacuum}). To this end we propose the following
change of coordinates and metric scaling
\begin{eqnarray}
t&=& \left( u-\Omega^2 v\right),\\
r&=& \left( u+\Omega^2 v\right),\\
\bar{g}_{\mu\nu}&=& \Omega^{-2}g_{\mu\nu}.
\end{eqnarray}
Because the asymptotic solution does not have any coupling dependence
the rescaling and redefinition are the usual ones \cite{blau1}.

Thus we obtain
\begin{equation}\label{ppAdS4}
d\bar{s}^2=\frac{8}{\alpha^2 u^2}\left( dudv+dx^2+dy^2 \right).
\end{equation}

Regarding the uplifting to eleven dimensions we get
\begin{eqnarray}\label{11ppAdS4}
\nonumber d\bar{s}^2&=&\frac{1}{\alpha^2}\Big[ \frac{8}{u^2}\left(
dudv+dx^2+dy^2 \right)+2d\xi^2+ \\ &&
\left(d\tilde{\rho}^2+d\tilde{\zeta}^2+d\tilde{\tau}^2+2d\tilde{\zeta}
d\tilde{\tau}\right)+\left(d\tilde{\rho}^2+d\tilde{\zeta}^2+d\tilde{\tau}^2+
2d\tilde{\zeta}d\tilde{\tau}\right)\Big].
\end{eqnarray}

In this form we obtain pp-wave configurations as limits of our
solutions with and without gauge fields.
Next we determine the amount of supersymmetry
preserved by these pp-wave configurations.
\subsection{$AdS_4$.}
It is straightforward to see this limit preserves all the
supersymmetries since the original $AdS_4$ solution does and what we
do corresponds to a renaming of coordinates.

\subsection{$AdS_2 \times H_2$}
From (\ref{eq:susy1}) and (\ref{eq:limitfield}) we see that the
condition $\epsilon_1,\epsilon_4=0$ of the near horizon supersymmetry
analysis no longer holds; furthermore
the equation (\ref{eq:susy2}) now reads
\begin{equation}
\delta\bar{\psi}_{\lambda}^{i}= \bar{\epsilon}^{i}
\overleftarrow{D}_{\lambda}-\alpha\bar{\epsilon}^{i}\gamma_{\lambda}=0.
\end{equation}
Hence
\begin{equation}
\bar{\epsilon}^{i}\left( \overleftarrow{\partial}_{\mu}+\frac{1}{2}\omega_{\mu ab}\sigma^{ab}-\alpha \gamma_{\mu}\right) +\bar{\epsilon}^{j}\left( 2\alpha A_{\mu}^{ij}\right)=0.
\end{equation}

For $i=1,4$ we can verify that
$\epsilon^1,\epsilon^4=\textrm{const}$ is always a solution of (\ref{eq:susy2}).
The non-zero spin connection for this
configuration is
\begin{equation}
\omega_{u\hat{u}\hat{v}}=\frac{1}{2u}.
\end{equation}

Plugging everything into (\ref{eq:susy2}) we obtain the following
equations
\begin{eqnarray}
\label{1ppAdS2}\bar{\epsilon}^2\left( \overleftarrow{\partial}_u+\frac{1}{2u}
\sigma^{\hat{p}\hat{q}}-\alpha \gamma_u \right)&=&0,\\
\bar{\epsilon}^2\left( \overleftarrow{\partial}_v-\alpha \gamma_v\right) &=&0,\\
\bar{\epsilon}^2(\overleftarrow{\partial}_p-\alpha\gamma_p)+\sqrt{2}\alpha k
\bar{\epsilon}^3&=&0,\\
\label{4ppAdS2}\bar{\epsilon}^2(\overleftarrow{\partial}_q-\alpha\gamma_q)+
\sqrt{2}\alpha k \bar{\epsilon}^3&=&0,
\end{eqnarray}
and similarly for $\epsilon^2 \leftrightarrow
\epsilon^3$. Let us observe that $\epsilon^{i}=\epsilon^{i}(u)$ solves
(\ref{1ppAdS2})-(\ref{4ppAdS2}), and
then we find that
$\epsilon^2\propto \epsilon^3$.

So we have that the pp-wave limit of the $AdS_2\times H_2$
enhances the supersymmetry to 3/4 of the total, which means there
are 24 supercharges conserved. As described in
\cite{super1,super2,super3} this corresponds to a supernumerary
supersymmetry.

\section{Discussion}
In this work we have obtained two supersymmetric solutions to
$D=4,N=4, SO(4)$ gauged supergravity, given by
(\ref{ansatz1},\ref{eh},\ref{ef}) and (\ref{AdS4}), which upon
uplifting become solutions of $D=11$ supergravity on $S^7$. Our
analysis was based on Cveti\^c et al embedding \cite{Cvetic1}
between these two theories. While one of the solutions is the
known $AdS_4$ vacuum, the other one corresponds to $AdS_2\times
H_2$. We refer to the former as asymptotic because it comes from
setting every matter field to zero which corresponds to their
spatial asymptotic behavior. As for the latter we call it near
horizon since near the $AdS_2$ horizon its contribution dominates
over the $H_2$. The supersymmetry preserved by the solutions is,
respectively, full (asymptotic) and 1/8 (near horizon) of the
original supersymmetry in four dimensions. We were able to find an
exact interpolating solution between the two.

In the case of $AdS_2\times H_2$ the IR flow can be obtained when
the wave length of the massless modes becomes large at low
energies compared with the size of the transverse space, thus
dropping all dependence in those coordinates. Thus, the
supergravity theory here studied is dual to a $2+1$ dimensional
Yang-Mills theory which in the low energy limit becomes
effectively 0+1 dimensional corresponding to a superconformal
Quantum Mechanics with 4 supercharges, a number not commonly
found.

Looking forward to study $AdS/CFT$ dualities of the kind of
theories here considered we derived the Penrose limit of our
$AdS_2\times H_2$ solution as given by (\ref{ppAdS2H2}). The
supersymmetry this limit preserves is 3/4 of the whole which is 24
supercharges, corresponding to a supernumerary type
\cite{super1,super2,super3}. Uplifting the $D=4$ pp-wave limit of
$AdS_2\times H_2$ (\ref{ppAdS2H2}) to eleven dimensions gave us
the $D=11$ pp-wave form (\ref{ppeleven}).

For completeness we have included the Penrose limit of the $AdS_4$ solution
(\ref{ppAdS4}) which is maximally supersymmetric. Its uplifting
to $D=11$ was performed giving (\ref{11ppAdS4}).

It would be of interest to check whether our analysis can be
extended in a more general context, for example, containing
families of supergravity solutions (see for instance
\cite{radu}), at least as general as the Cveti\^c et al analysis
allows to do. Furthermore, either in our case or a more general
one looking for the dual SYM theory might shed light on the
$AdS/CFT$ correspondence. We leave such analysis for future work.

\section*{Acknowledgments}
It is a pleasure to thank Carlos Nu\~nez for his
patience and advise all the way in the course of this work. Also
we are grateful to
 Juan Maldacena for illuminating comments.
This work was partially supported by CONACyT-NSF E120-0837 grant as well as 
CONACyT-32431-E. The work of H. Hern\'andez was supported by CONACyT scholarship
No. 113375. He is also grateful to ICTP for support under the Young Collaborator Programme.


\newpage
\begin{thebibliography}{99}
\bibitem{malda1} J. Maldacena, \emph{The large N-limit
 of superconformal field theories and supergravity}, Adv. Theor. Math. Phys. 2
 (1998)231, [hep-th/971120].
\bibitem{gubser} S. S. Gubser, R. Klebanov, A. M. Polyakov,
 \emph{Gauge theory correlators from noncritical string theory},
 Phys. Lett. B428 (1998) 105, [hep-th/9802109].
\bibitem{witten} E. Witten, \emph{AdS space and holography}, Adv. Theor. Math.
Phys.2 (1998)253, [hep-th/9802150].
\bibitem{edelstein} J. D. Edelstein, A. Paredes, A. V. Ramallo,
 \emph{Let's twist again: general metrics of $G(2)$ holonomy from
 gauged supergravity}, JHEP 0301(2003)011, [hep-th/0211203].
\bibitem{malda2} N. Itzhaki, J. Maldacena, J. Sonnenschein, S. Yankielowicz,
 \emph{Supergravity and the large-N limit of theories with sixteen supercharges},
 Phys. Rev. D58 (1998)46004, [hep-th/9802042].
\bibitem{super1} M. Cveti\^c, H. L\"u, C. Pope, \emph{M theory PP
 waves, Penrose limits and supernumerary supersymmetries},
 Nucl. Phys. B644(2002)65-84, [hep-th/0203229].
\bibitem{super2} M. Cveti\^c, H. L\"u, C. Pope, \emph{Penrose limits, PPwaves
 and deformed M2 branes}, [hep-th/0203082].
\bibitem{Nunez1} J. Maldacena, C. Nu\~nez, \emph{Supergravity description of
 field theories on curved manifolds and a no go theorem},
 Int. J. Mod. Phys. A16(2001)822-855, [hep-th/0007018].
\bibitem{Cvetic1} M. Cveti\^c, H. L\"u ,C.N. Pope,
 \emph{Four dimensional $N=4$, $SO(4)$ gauged supergravity from $D=11$},
 Nuc. Phys. B574(2000) 761-781, [hep-th/9910252].
\bibitem{Cvetic2} M. Cveti\^c, M. Duff, P. Hoxha, L. Liu, H. L\"u, R. Acosta, C. Pope, H. Sati,
 T. Tran, \emph{Embedding AdS black holes in ten and eleven
 dimensions}, Nucl. Phys. B558(1999)96-126, [hep-th/9903214].
\bibitem{Cvetic3} M. Cveti\^c, H. L\"u, C. Pope, \emph{Consistent
 Kaluza-Klein sphere reductions}, Phys. Rev. D62(2000)064028, [hep-th/0003286].
\bibitem{stromin1} J. Maldacena, J. Michelson, A. Strominger,
 \emph{Anti-De Sitter fragmentation}, JHEP 9902(1999)011 , [hep-th/9812073].
\bibitem{stromin2} J. Michelson, A. Strominger, \emph{The geometry of (super) conformal
 quantum mechanics}, Commun. Math. Phys. 213(2000)1-17, [hep-th/9907191].
\bibitem{stromin3} J. Michelson, A. Strominger, \emph{Superconformal multiblack
 hole quantum mechanics}, JHEP 9909(1999)005, [hep-th/9908044].
\bibitem{sols1} M. Schvellinger, T. Tran, \emph{Supergravity duals
 of gauge field theories from SU(2)$\times$U(1)
 gauged supergravity in five dimensions}, JHEP 0106(2001)25,
 [hep-th/0105019]. J. Maldacena, H. Nastase, \emph{The supergravity
 dual of a theory with dynamical supersymmetry breaking}, JHEP
 0109(2001)24, [hep-th/0105049]. J. P. Gauntlett, N. Kim, S. Pakis,
 D. Waldram, \emph{Membranes wrapped on holomorphyc curves}, Phys.
 Rev. D65(2002)26003, [hep-th/0105250]. R. Hern\'andez, \emph{Branes
 wrapped on coassociative cycles}, Phys. Lett. B521(2001)371,
 [hep-th/0106055]. J. Gauntlett, N. Kim, D. Martelli, D. Waldram,
 \emph{Wrapped fivebranes and $N=2$ super Yang-Mills theory}, Phys.
 Rev. D64(2001)106008, [hep-th/0106117]. F. Bigazzi, A. Cotrone, A.
 Zaffaroni, \emph{$N=2$ gauge theories from wrapped fivebranes},
 Phys. Lett. B519(2001)269, [hep-th/0106160]. J. Gomis, T. Mateos, \emph{D6 branes
 wrapping Kaehler four-cycles}, Phys. Lett. B524(2002)170, [hep-th/0108080]. J.
 Gauntlett, N. Kim, \emph{M-fivebranes wrapped in supersymmetric
 cycles}, Phys. Rev. D65(2002)86003 [hep-th/0012195]. J. Gomis, J. Russo,
 \emph{$D=2+1,N=2$ Yang-Mills theory from wrapped branes},
 JHEP 0110(2001)028, [hep-th/0109177]. J. Gauntlett, N. Kim, D. Martelli, D. Waldram,
 \emph{Fivebranes wrapped on SLAG three-cycles and related
 geometry}, JHEP 0111(2001)18, [hep-th/0110034]. J. Gomis, \emph{On SUSY breaking and
 chiSB from string duals}, Nuc. Phys. B624(2002)181, [hep-th/0111060]. P. DiVecchia,
 H. Enger, E. Imeroni, E. Lozano-Tellechea, \emph{Gauge theory from
 wrapped and fractional branes}, Nuc. Phys. B631(2002)95, [hep-th/0112126].
\bibitem{berenstein} D. Berenstein, J. Maldacena, H. Nastase, \emph{Strings in flat space
 and pp waves from ${\cal N}=4$ Super Yang Mills}, JHEP 0204
 (2002)03, [hep-th/0202021]
\bibitem{penrose} R. Penrose in \emph{Differential geometry and relativity}, ed. M. Cahen and M. Flato, Reidel, Netherlands 1976.
\bibitem{guven} R. G\"uven, Phys. Lett. B482 (2000) 255.
\bibitem{correa1} D.H. Correa, E.F. Moreno, S. Reuillon, F.A.
 Schaposnik, \emph{PP-waves from BPS supergravity monopoles},
 Phys. Lett. B552(2003)280-286, [hep-th/0211093].
\bibitem{nunez2} C. Nu\~nez, I. Park, M. Schvellinger, T. Tran,
 \emph{Supergravity duals of gauge theories from $F(4)$ gauged supergravity in six dimensions},
 JHEP 04(2001) 25, [hep-th/0103080].
\bibitem{nastase} D. Berenstein, J. Maldacena, H. Nastase,
 \emph{Strings in flat space and pp waves from $N=4$ super
 Yang-Mills}, JHEP 0204(2002)013, [hep-th/0202021].
\bibitem{super3} J. Gauntlett, C. Hull, \emph{PP-waves in 11 dimensions
 with extra supersymmetry}, JHEP 0206(2002)013, [hep-th/0203255].
\bibitem{Das1} A. Das, \emph{$SO(4)$-invariant extended supergravity},
 Phys. Rev. D15(1977)2805.
\bibitem{freedman} C. Burges, D. Freedman, S. Davis, W. Gibbons,
 \emph{Supersymmetry in AntideSitter space}, Ann.
  Phys.167(1986)285.
\bibitem{lu1} H. L\"u, C. Pope, P. K. Townsend, \emph{Domain walls
 from Anti-de Sitter spacetime}, Phys. Lett. B391(1997)39-46, [hep-th/9607164].
\bibitem{lu2} H. L\"u, C. Pope, J. Rahmfeld, \emph{A construction of
 Killing spinors on $S^n$}, J. Math. Phys. 40(1999)4518-4526, [hep-th/9805151].
\bibitem{Romans1} L. J. Romans, \emph{Supersymmetric, cold and
 lukewarm black holes in cosmological Einstein-Maxwell theory},
 Nucl. Phys. B383(1992)395-415,  [hep-th/9203018].
\bibitem{blau2} M. Blau, \emph{Killing spinors and SYM on curved spaces},
 JHEP 0011(2000)023, [hep-th/0005098].
\bibitem{Das2} A. Das, M. Fischler, \emph{Super-Higgs effect
 in a new class of scalar models and a model of super QCD}, Phys. Rev. D16 (1977)3427.
\bibitem{Barton1} S. Gates, B. Zweibach,
 \emph{Gauged $N=4$ supergravity theory with a new scalar potential},
 Phys. Lett. 123B(1983)200.
\bibitem{barton2} J. Gates, B. Zweibach, \emph{Searching for
 all $N=4$ supergravities with global $SO(4)$}, Nuc. Phys. B238 (1984)99.
\bibitem{sabra} W. A. Sabra, \emph{Curved branes in AdS Einstein-Maxwell gravity and Killing
 spinors}, Phys. Lett. B552(2003)247-254, [hep-th/0210265].
\bibitem{blau1} M. Blau, J. Figueroa-O'Farril, G. Papadopoulos,
 \emph{Penrose limits and maximal supergravity},
 Class. Quant. Grav. 19(2002)L87-L95 , [hep-th/0201081].
\bibitem{fuji} H. Fuji, K. Ito, Y. Sekino, \emph{Penrose limit and
 string theories on various brane backgrounds}, JHEP 0211(2002)005, [hep-th/0209004].
\bibitem{guven2} R. G\"uven, \emph{Plane wave limits and T-duality},
 Phys. Lett. B482(2000)255-263, [hep-th/0005061].
 \bibitem{radu} E. Radu, \emph{New nonabelian solution in D=4, N=4 gauged
 supergravity}, Phys. Lett. B542(2002)225-281, [gr-qc/0202103].
\end{thebibliography}
\end{document}